\documentclass[conference]{IEEEtran}
\IEEEoverridecommandlockouts
\usepackage{cite}
\usepackage{amsmath,amssymb,amsfonts}
\usepackage{algorithmic}
\usepackage{graphicx}
\usepackage{textcomp}
\usepackage{xcolor}
\usepackage{CJKutf8}

\def\BibTeX{{\rm B\kern-.05em{\sc i\kern-.025em b}\kern-.08em
		T\kern-.1667em\lower.7ex\hbox{E}\kern-.125emX}}

\begin{document}
\begin{CJK*}{UTF8}{gbsn}
	
\title{Ensemble Methodology: Innovations in Credit Default Prediction Using LightGBM, XGBoost, and LocalEnsemble
}

\author{\IEEEauthorblockN{1\textsuperscript{st} Mengran Zhu}
\IEEEauthorblockA{
\textit{Miami University}\\
OH, USA \\
mengran.zhu0504@gmail.com}
\and
\IEEEauthorblockN{2\textsuperscript{nd} Ye Zhang}
\IEEEauthorblockA{\textit{University of Pittsburgh} \\
Pittsburgh, USA \\
yez12@pitt.edu}
\and
\IEEEauthorblockN{3\textsuperscript{rd} Yulu Gong}
\IEEEauthorblockA{\textit{Northern Arizona University}\\
AZ, USA \\
yg486@nau.edu}
\and
\IEEEauthorblockN{4\textsuperscript{th} Kaijuan Xing}
\IEEEauthorblockA{\textit{University of Texas at Austin}\\
Austin, USA \\
xingkj@utexas.edu}
\and
\IEEEauthorblockN{5\textsuperscript{th} Xu Yan}
\IEEEauthorblockA{\textit{Trine University}\\
Indiana, USA \\
tianbosong@aol.com}
\and
\IEEEauthorblockN{6\textsuperscript{th} Jintong Song}
\IEEEauthorblockA{\textit{Boston University}\\
Boston, USA \\
jintongs@bu.edu}
}

	\maketitle
	
	\begin{abstract}
        In the realm of consumer lending, accurate credit default prediction stands as a critical element in risk mitigation and lending decision optimization. Extensive research has sought continuous improvement in existing models to enhance customer experiences and ensure the sound economic functioning of lending institutions. This study responds to the evolving landscape of credit default prediction, challenging conventional models and introducing innovative approaches. By building upon foundational research and recent innovations, our work aims to redefine the standards of accuracy in credit default prediction, setting a new benchmark for the industry.

        To overcome these challenges, we present an Ensemble Methods framework comprising LightGBM, XGBoost, and LocalEnsemble modules, each making unique contributions to amplify diversity and improve generalization. By utilizing distinct feature sets, our methodology directly tackles limitations identified in previous studies, with the overarching goal of establishing a novel standard for credit default prediction accuracy. Our experimental findings validate the effectiveness of the ensemble model on the dataset, signifying substantial contributions to the field. This innovative approach not only addresses existing obstacles but also sets a precedent for advancing the accuracy and robustness of credit default prediction models.
        
	\end{abstract}
	
	
	\section{Introduction}
    
        In the consumer lending landscape, precise credit default prediction is vital for risk mitigation and optimal lending decisions. The American Express - Default Prediction competition seeks innovative machine learning solutions to surpass industry models. As shown in Figure 1, the task involves predicting future credit card default by assessing a customer's historical monthly profile. The binary target variable, "default" or "non-default," depends on whether the customer pays the entire outstanding credit card balance within 120 days after the billing date. The substantial test dataset includes 900,000 customers, 11 million records, and 191 variables.

        \begin{figure}[htbp]
        \centering
        \includegraphics[width=1\linewidth]{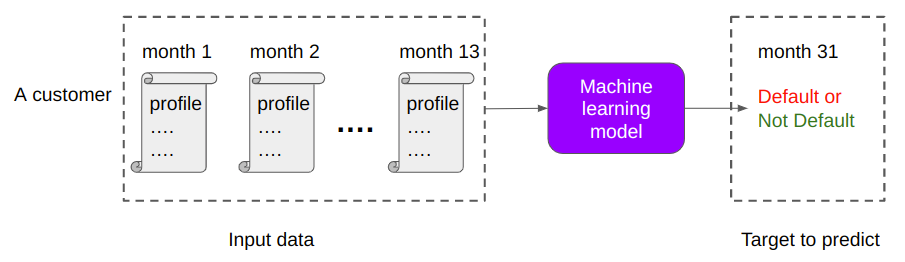}
        \caption{Problem Overview Credit Default Prediction}
        \label{fig:problem-overview-credit-default-prediction}
        \end{figure}

        Early credit default prediction research, exemplified by studies like Sayjadah et al.\cite{sayjadah2018credit} and Soui et al.\cite{soui2018credit}, utilized machine learning but lacked detailed outcomes. Subsequent works, including Yang and Zhang's data mining analysis\cite{yang2018comparison}, Yu's exploration of imbalanced datasets\cite{yu2020application}, and Alam et al.'s study\cite{alam2020investigation}, offered insights with limitations. Chen and Zhang\cite{chen2021research} introduced a unique perspective but indicated a need for further advancements in the field. 
        
        Recent studies, like Gao et al.'s XGBoost-LSTM\cite{gao2021research} and Zheng's fusion of XGBoost and LightGBM\cite{zheng2022default}, bring innovation to credit default prediction. While contributing to the evolving landscape, these works highlight areas for improvement. Guo et al.\cite{guo2023credit} and Gan et al.\cite{gan2023lightgbm} explore ensemble models, showing promise but leaving space for methodological enhancements. These endeavors collectively underscore the dynamic nature of credit default prediction research and the continuous quest for more effective models.

        In response to the existing gaps and challenges in credit default prediction, our approach revolves around Ensemble Methods\cite{dietterich2000ensemble}. We have meticulously designed three key modules: the LightGBM module\cite{ke2017lightgbm}, the XGBoost module\cite{chen2016xgboost}, and the LocalEnsemble\cite{miyoshi2007local}. Each module plays a distinct role in the final prediction, leveraging different sets of features to enhance diversity and generalize better. The LightGBM and XGBoost modules bring the strengths of these individual models, while the LocalEnsemble module focuses on integrating local predictions for improved overall accuracy. This ensemble approach aims not only to challenge the current models in production but also to set a new benchmark for credit default prediction, offering a robust and comprehensive solution.

        The main contributions of this work can be summarized as follows:

        \begin{itemize}
            \item We propose a novel Ensemble Methods framework comprising three key modules – LightGBM, XGBoost, and LocalEnsemble. This strategic integration harnesses the strengths of individual models, enhancing diversity and generalization for more robust credit default predictions.
            \item  To capture the interactions between various influencing factors more accurately, we design a Local Ensemble module. This module models different models using distinct feature combinations to enhance diversity and improve generalization.
            \item Our experiments confirm the efficacy of our Ensemble Model on the American Express dataset.
        \end{itemize}
        
	\section{Related Work}

        In the early stages of delving into credit default prediction, researchers made valuable contributions by employing machine learning techniques and framing the issue as a classification challenge. Sayjadah et al.\cite{sayjadah2018credit} concentrated on applying diverse algorithms, lacking explicit details on outcomes. Meanwhile, Soui et al.\cite{soui2018credit} addressed it as a classification problem without specifying the algorithms used or presenting performance metrics. Both studies played a role in the foundational exploration of credit default prediction, paving the way for subsequent research in this domain.

        Yang and Zhang\cite{yang2018comparison} performed an extensive analysis, comparing various data mining methods in the context of credit card default prediction. Their study yielded valuable insights into the comparative performance of different techniques. Building upon this, Yu\cite{yu2020application} and Alam et al. \cite{alam2020investigation} delved deeper into the application of machine learning algorithms, with a specific focus on exploring prediction models in imbalanced datasets. Concurrently, Chen and Zhang\cite{chen2021research} introduced a distinctive perspective by combining k-means SMOTE and BP neural networks to enhance prediction accuracy. Collectively, these studies contribute to the evolving landscape of credit default prediction, offering diverse methodologies and insights that enrich the understanding and potential improvements in this critical domain.

        Recent investigations in credit default prediction showcase innovative strategies, exemplified by Gao et al.'s utilization of XGBoost-LSTM\cite{gao2021research} and Zheng's integration of both XGBoost and LightGBM\cite{zheng2022default}. Moreover, Guo et al.\cite{guo2023credit} explored ensemble models applied to time-series behavioral data, expanding the methodological spectrum. Simultaneously, Gan et al.\cite{gan2023lightgbm} narrowed their focus to a LightGBM-based model customized for American Express. These diverse methodologies contribute to the evolution of credit default prediction models, providing a comprehensive array of techniques. Each study offers unique insights, collectively enriching the understanding of effective approaches and potential advancements in predicting credit defaults.
        
	\section{Methodology}
 
            \subsection{Data Preprocessing}
        The primary objective of data preprocessing is to eliminate missing values and outliers from the original dataset, filter features relevant to the learning task, and transform them into a format acceptable for model input and computation. As shown in Figure 2, our data preprocessing involves the following steps:
        
        \textbf{Noise Removal:} The data is artificially injected with random uniform noise. To rectify this, a rounding method is employed for denoising. This step ensures the accuracy and quality of the data, laying a solid foundation for subsequent analysis and model training.
        
        \textbf{Type Conversion:} The original dataset comprises 188 features at an industrial scale, all of which are of floating-point type. Hence, the dataset must be converted to int8/int16/float32 for computational purposes.

        \textbf{Outlier Handling:} Outliers are filtered out by setting the attributes of exceptional records to NaN, different feature engineering approaches may employ various methods for filling missing data.
            
        \begin{figure}[htbp]
        \centering
        \includegraphics[width=1\linewidth]{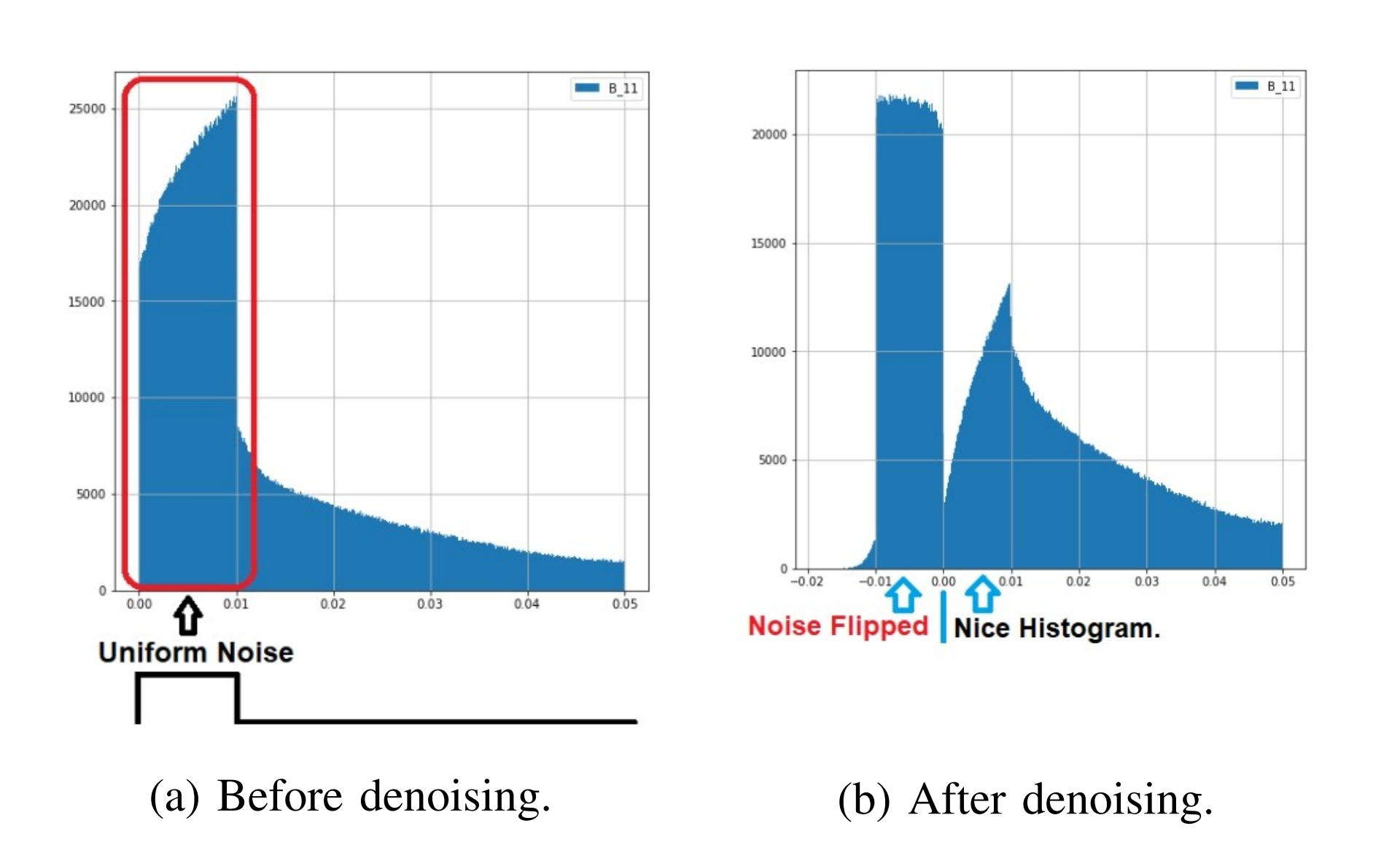}
        \caption{Data Preprocessing}
        \label{fig:Data Preprocessing}
        \end{figure}

            \subsection{Feature Engineering}
    
        In the training set, we have data for 458,913 customers, while the test set contains information for 924,621 customers. Among these customers, 80\% have complete records of 13 statements, while the remaining 20\% have between 1 and 12 statements. Due to the anonymity of attribute labels, we adopt a series of feature engineering steps to precisely describe and reveal potential feature relationships:
        \begin{itemize}
          \item \textbf{Aggregated Features:} We compute aggregated features for each user based on the 13 statements, reflecting the user's situation from different perspectives.
            \begin{itemize}
              \item Continuous values: ['mean', 'std', 'min', 'max', 'last', 'median']
              \item Discrete values: ['count', 'last', 'nunique']
            \end{itemize}
          \item \textbf{Lag Features:} Recent changes in a user's statement may lead to defaults. Lag features capture the difference between 'last' and 'mean' to reflect whether the user has experienced recent changes in behavior patterns.
          \item \textbf{Meta Features:} As shown in Figure 3, Initially, we merge the target variable with the training set based on customer\_id. This integration ensures that each data point in the set has a corresponding target variable linked to the customer\_id.
    
            \begin{figure}[htbp]
            \centering
            \includegraphics[width=0.6\linewidth]{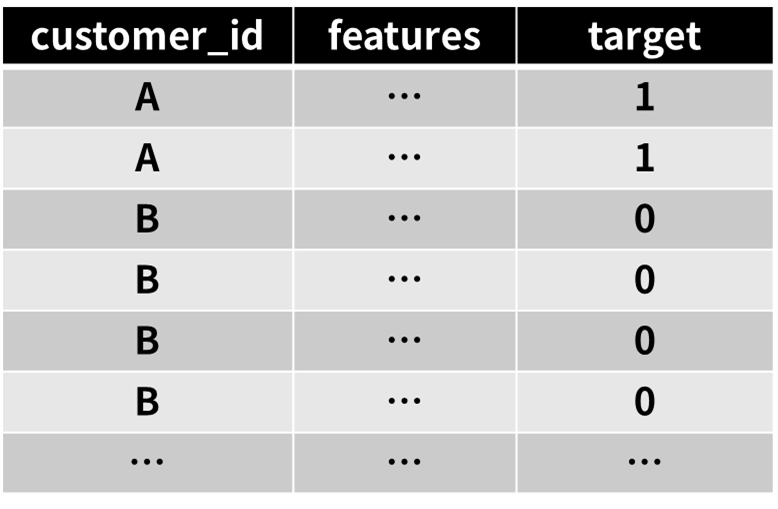}
            \caption{Feature Engineering Step One}
            \label{fig:Feature Engineering Step One}
            \end{figure}
            
          As shown in Figure 4, Utilizing the engineered features as inputs and the target as outputs, we train a tree model, such as LightGBM, incorporating k-fold cross-validation. The out-of-fold (OOF) predictions obtained during this process serve as meta features. These predictions are then directly incorporated as 13 continuous value features in subsequent model training.

        \begin{figure}[htbp]
        \centering
        \includegraphics[width=0.6\linewidth]{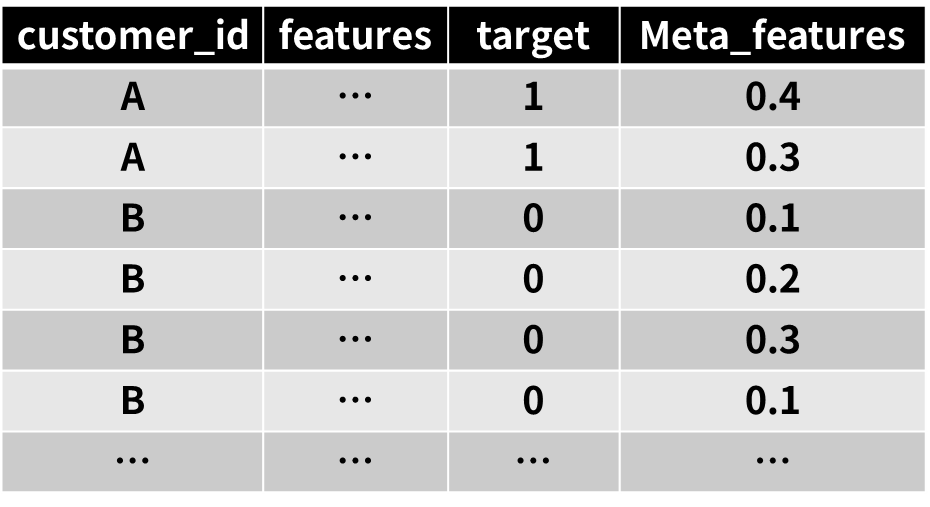}
        \caption{Feature Engineering Step Two}
        \label{fig:Feature Engineering Step Two}
        \end{figure}
        
        \end{itemize}
    
        \subsection{Ensemble Model }
        In this section, we will introduce our ensemble model consisting of three primary modules: the LightGBM module\cite{ke2017lightgbm}, the XGBoost module\cite{chen2016xgboost}, and the LocalEnsemble\cite{miyoshi2007local}. As shown in Figure 5, each module plays a distinct role in the final prediction, and we elaborate on the details of each module and the fusion method below. To achieve better integration results, we employ different feature sets for each model to enhance diversity and improve generalization.Mathematically, this is represented as:
    
            \begin{equation}
            \hat{y}_e = \sum_{i=1}^{N} w_i \cdot \hat{y}_i 
            \end{equation}
    
        Where $\hat{y}_e$ represents the ensemble prediction, $\hat{y}_i$ represents the prediction from the $i$-th model, and $w_i$ represents the weight assigned to the $i$-th model. 
    
            \begin{figure}[htbp]
            \centering
            \includegraphics[width=1\linewidth]{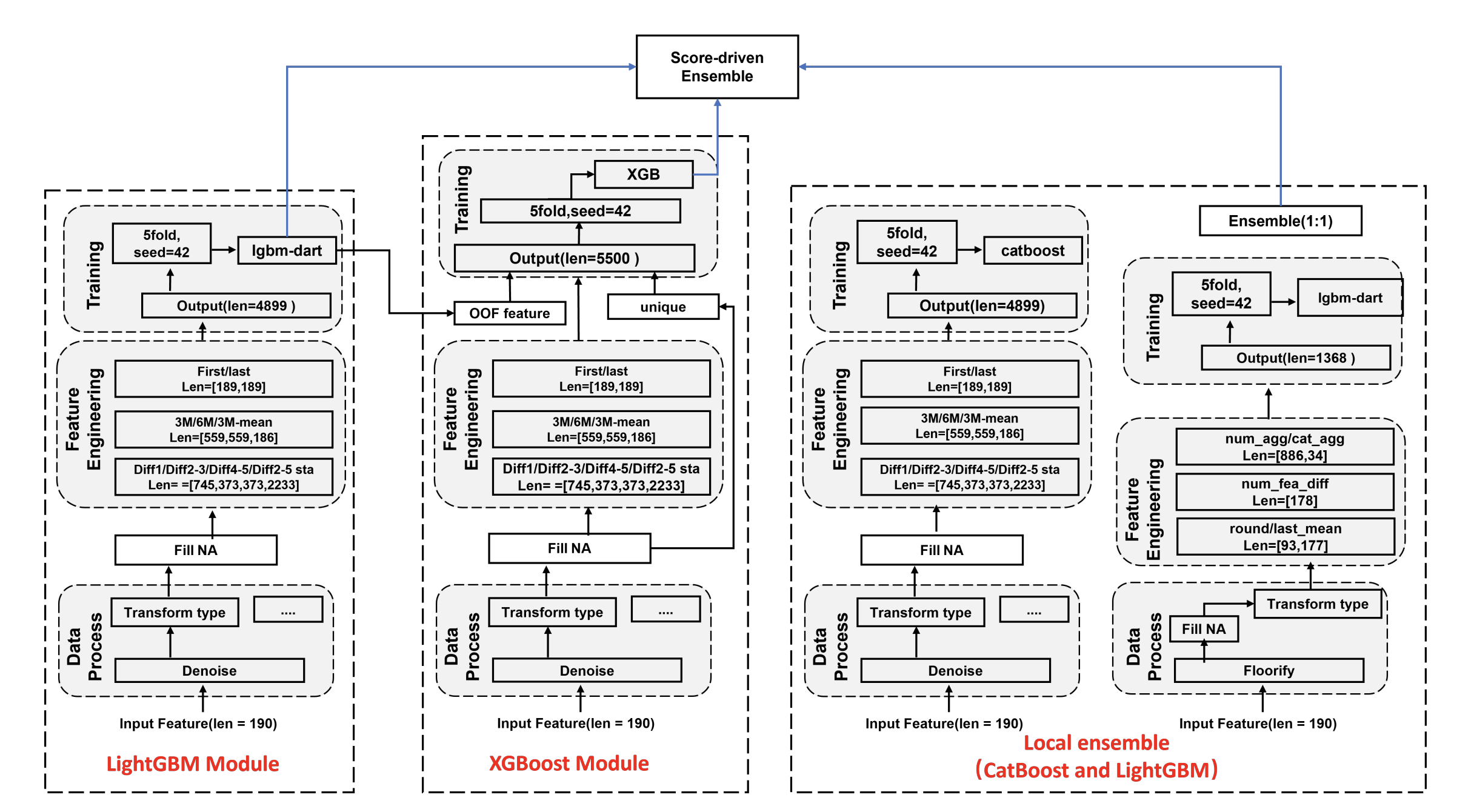}
            \caption{Ensemble Model}
            \label{fig:Ensemble Model}
            \end{figure}
            
        \subsubsection{LightGBM Module}
        LightGBM\cite{ke2017lightgbm}, an efficient gradient boosting decision tree, focuses on reducing training data points while retaining information through the GOSS algorithm. The preprocessing involves feature engineering, considering the most recent 6 months of behavioral records for 80\% of customers. Aggregation operations are applied to generate statistical features, enhancing the relationship between historical time points and current behavior.
    
        \subsubsection{XGBoost Module}
        XGBoost\cite{chen2016xgboost}, an optimized distributed gradient boosting library, employs a broader range of feature engineering data compared to the LightGBM module. It utilizes unique features and the out-of-fold (OOF) technique. Categorical features are handled using independent encoding or one-hot encoding. The OOF feature integrates predictions from the LightGBM module, serving as a foundational model, while the XGBoost module acts as the meta-model.
        
        \subsubsection{Local Ensemble}
        To capture the interactions between various influencing factors more accurately, we design a Local Ensemble module\cite{miyoshi2007local}. This module models different models using distinct feature combinations to enhance diversity and improve generalization.
        
        \begin{itemize}
        \item \textbf{CatBoost(Local)}\cite{prokhorenkova2018catboost}, a GBDT framework based on symmetric decision trees, focuses on effective handling of categorical features. Similar to the LightGBM module, CatBoost undergoes the same data processing and feature engineering. It is employed to describe the impact and effectiveness of different models on the same feature set.
        
        \item \textbf{LightGBM(Local)}, an approach that employs distinct data preprocessing and feature engineering methods, utilizing an entirely different dataset and feature combination. This enhances the overall ensemble model's generalization capability, mitigating overfitting issues observed with a single dataset. However, its performance is not as effective as the LightGBM module.
        \end{itemize}

            \section{Experiments}
        \subsection{Datasets}
        The credit default prediction dataset aims to forecast the probability of a customer defaulting on their credit card balance. The binary target variable is determined within an 18-month window after the latest statement, marking default if the due amount remains unpaid within 120 days. Features, anonymized and normalized, include categories like Delinquency, Spend, Payment, Balance, and Risk variables, with additional categorical features.

        To ensure class balance, the dataset underwent a 5\% subsampling of the negative class. It includes anonymized customer profiles and time-series behavioral data. The training set features a binary target variable based on payment behavior within an 18-month window, while the testing set predicts future credit defaults using monthly customer profiles. Diverse feature categories offer insights for effective credit default prediction models in consumer lending.
                
        \subsection{Evalution Metrics}

        The evaluation metric (\(M\)) for this credit default prediction competition combines two rank-ordering measures: the Normalized Gini Coefficient (\(G\))\cite{raffinetti2015gini} and the default rate captured at 4\% (\(D\)).

        \begin{equation}
        M = 0.5 \cdot (G + D)
        \end{equation}

        \textbf{Normalized Gini Coefficient (\(G\)):}
        \begin{itemize}
          \item Measures the discriminatory power of the model's predicted probabilities.
          \item Represents the area between the Receiver Operating Characteristic (ROC) curve and the diagonal line, normalized to the maximum possible area.
        \end{itemize}
        
        \textbf{Default Rate Captured at 4\% (\(D\)):}
        \begin{itemize}
          \item Indicates the percentage of positive labels (defaults) captured within the highest-ranked 4\% of predictions.
          \item Serves as a Sensitivity/Recall statistic.
        \end{itemize}
        
        Both \(G\) and \(D\) sub-metrics assign a weight of 20 to negative labels to account for downsampling. The combined metric \(M\) ranges from 0 to 1, with a higher value indicating superior model performance. The evaluation metric provides a comprehensive assessment, considering both discriminatory power and sensitivity in predicting credit defaults.
            
        As shown in Figure 6, this metric has a graphical interpretation, optimizing both the area under the red curve and the intersection point with the green line. Illustrated in Figure 5, the normalized Gini coefficient, stretched to obtain the Area Under the Curve (AUC), is represented by the shaded area under the red curve (0 to 1 range). The normalized Gini coefficient is calculated as \(2 \times \text{AUC} - 1\), ranging from -1 to 1. A larger red area indicates better model performance. Capturing the 4\% default rate is achieved by setting the threshold at 4\% of the total weighted sample count, resulting in the True Positive Rate or Recall (0 to 1). This corresponds to the coordinate position at the intersection of the green line and the red ROC curve. A higher intersection point indicates superior model performance.

            \begin{figure}[htbp]
            \centering
            \includegraphics[width=0.8\linewidth]{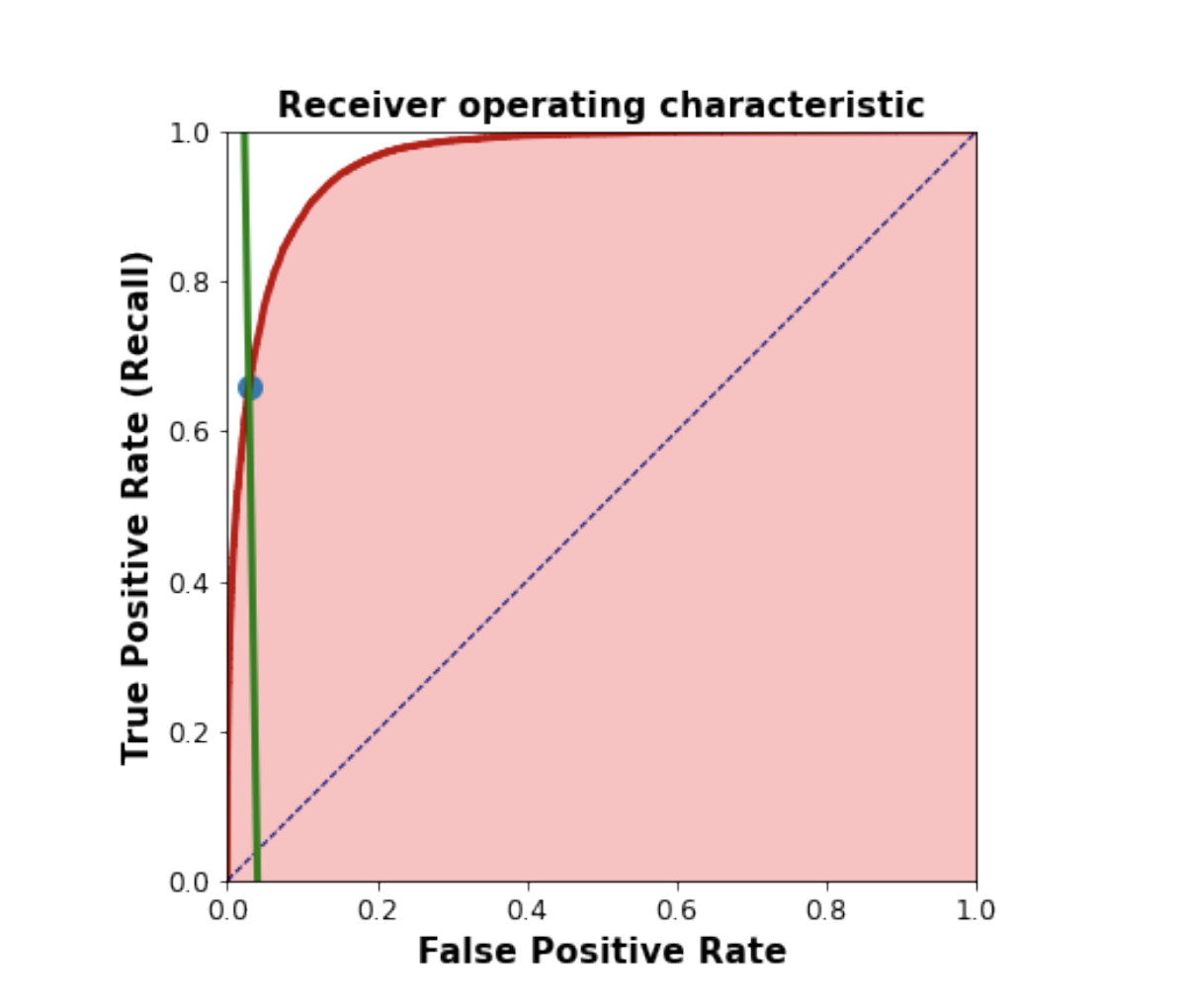}
            \caption{Evaluation Metrics Graphical Interpretation}
            \label{fig:Evaluation_Metrics} 
            \end{figure}

            \subsection{Results}
        In this section, we compared the performance of various models in the American Express - Default Prediction competition. The evaluated models include deep learning based model: GRU\cite{dey2017gate}, Transformer\cite{vaswani2017attention}, Tabtransformer\cite{huang2020tabtransformer}, and Neural Networks\cite{abdi1999neural}. Machine learning based model: XGBoost\cite{chen2016xgboost}, LightGBM\cite{ke2017lightgbm}, CatBoost (Local)\cite{prokhorenkova2018catboost}, LightGBM (Local), Local Ensemble\cite{miyoshi2007local}, and our proposed Ensemble Model. Each model was assessed based on the public and private datasets, with scores provided in Table 1: 
        \begin{table}[htbp]
        \caption{Model Performance Comparison}
        \centering
        \renewcommand{\arraystretch}{1.2} 
        \setlength{\tabcolsep}{6pt} 
        \begin{tabular}{|c|c|c|}
        \hline
        \textbf{Model} & \textbf{Public Score (49\%)} & \textbf{Private Score (51\%)} \\
        \hline
        GRU & 0.78877 & 0.79832 \\
                \hline
        Transformer & 0.78916 & 0.79832 \\
                \hline
        Tabtransformer & 0.78271 & 0.79236 \\
                \hline
        Neural Networks & 0.78705 & 0.79698 \\
                \hline
        XGBoost & 0.79982 & 0.80757 \\
                \hline
        LightGBM & 0.80006 & 0.80809 \\
                \hline
        CatBoost (Local) & 0.79804 & 0.80629 \\
                \hline
        LightGBM (Local) & 0.79967 & 0.80697 \\
                \hline
        Local Ensemble & 0.80094 & 0.80842 \\
                \hline
        Ensemble Model & \textbf{0.80128} & \textbf{0.80872} \\
        \hline
        \end{tabular}
        \label{tab:results}
        \end{table}
        
        Our Ensemble Model outperformed others in both public and private datasets, attaining the highest scores. This highlights the strategic integration of LightGBM, XGBoost, and LocalEnsemble modules, leveraging individual strengths for diversity and generalization. The model sets a new benchmark, offering a comprehensive solution adaptable to diverse scenarios.

        \subsection{Features Importance Analysis}

        To gain a deeper understanding of the role each feature plays in the models, we employed specific methods to calculate and visualize feature importance. In the XGBoost model, we chose Average Gain as the primary criterion, measuring the average performance improvement each feature brings in tree node splits. For the LightGBM model, we utilized Total Information Gain, representing the sum of information gain contributed by a feature across all splitting nodes.

            \begin{figure}[htbp]
            \centering
            \includegraphics[width=1\linewidth]{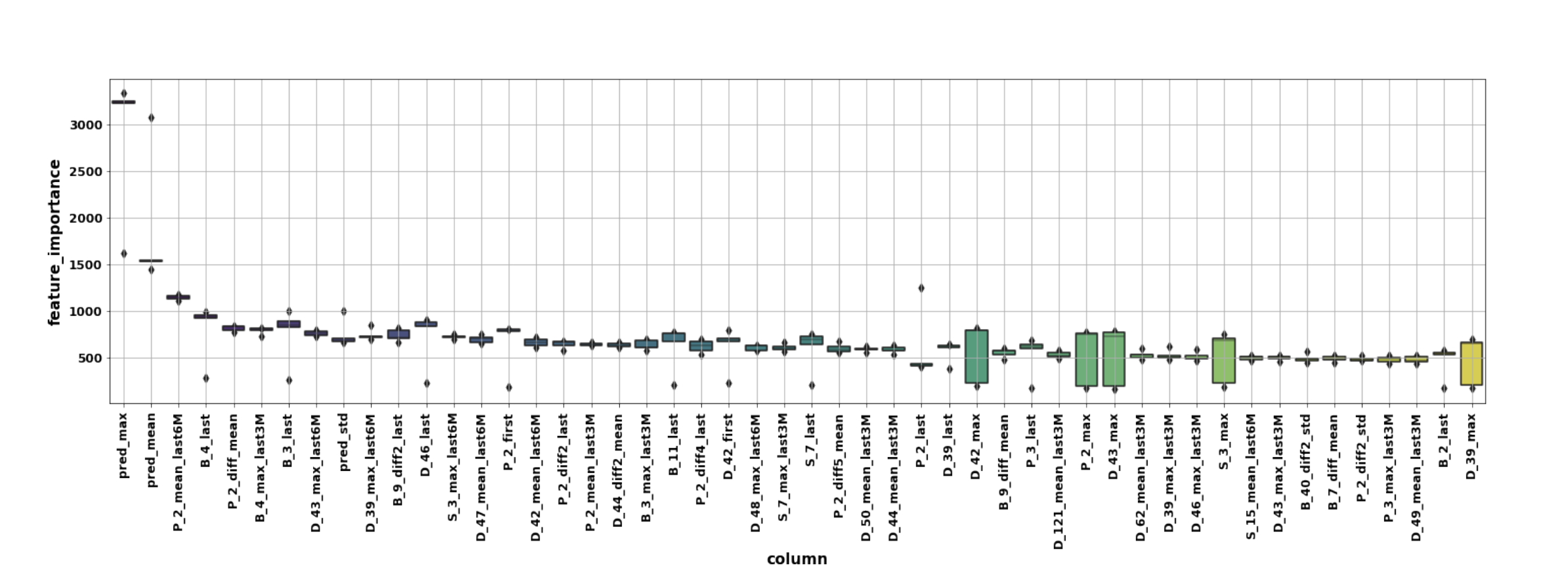}
            \caption{Features Importance of Xgboost}
            \label{fig:Features Importance of Xgboost}
            \end{figure}

            \begin{figure}[htbp]
            \centering
            \includegraphics[width=1\linewidth]{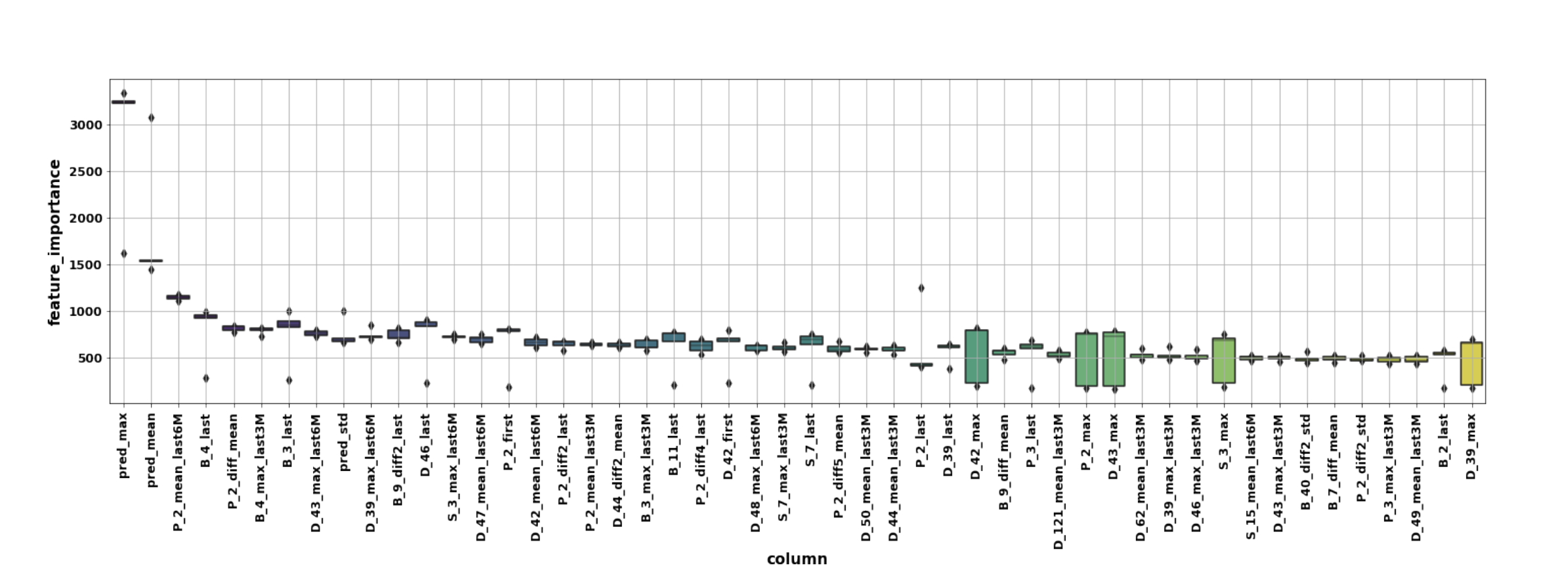}
            \caption{Features Importance of Lightgbm}
            \label{fig:Features Importance of Lightgbm}
            \end{figure}

        As shown in Figure 7 and Figure 8, through 5-fold cross-validation, we computed feature importance for each fold, presenting overall importance using Box-plots. Averaging results, the XGBoost model, with 5500 features, showed the top 50 features contributing over 90\%. A parallel trend was observed in the LightGBM model, highlighting the effectiveness of prioritizing key features in predicting credit defaults, ensuring model interpretability and practical applicability.
         
            \section{Conclusion}
        In conclusion, this study tackles the crucial issue of predicting credit defaults in consumer lending. While subsequent works provided valuable insights, they also had limitations, indicating the necessity for further advancements. Recent studies, such as those employing innovative approaches like XGBoost-LSTM and fusion methods, have contributed to the evolving landscape, shedding light on areas that can be enhanced.

        To address existing gaps and challenges, our proposed Ensemble Methods framework, which includes LightGBM, XGBoost, and a LocalEnsemble module, aims to establish a new benchmark for credit default prediction. Each module contributes uniquely by leveraging diverse feature sets, enhancing model diversity, and improving generalization. The Ensemble Model, integrating the strengths of individual modules, demonstrates efficacy on the American Express dataset, presenting a robust and comprehensive solution. Additionally, our design of the Local Ensemble module enhances accuracy by modeling different combinations of features, providing a strategic response to the ongoing pursuit of more effective credit default prediction models. Our experiments affirm the efficacy of the Ensemble Model on the American Express dataset.
        \bibliographystyle{IEEEtran}
        \bibliography{references}
\end{CJK*}	
\end{document}